\documentclass[twocolumn]{aastex61}

\submitjournal{ApJ}

\shorttitle{Detecting Gravitational Waves with Astrometry}
\shortauthors{Darling, Truebenbach, \& Paine}

\begin{document}

\title{Astrometric Limits on the Stochastic Gravitational Wave Background}

\correspondingauthor{Jeremy Darling}
\email{jeremy.darling@colorado.edu}

\author[0000-0003-2511-2060]{Jeremy Darling}
\affil{Center for Astrophysics and Space Astronomy, 
Department of Astrophysical and Planetary Sciences,
University of Colorado, 389 UCB,
Boulder, CO 80309-0389, USA}
\author{Alexandra E. Truebenbach}
\affil{Center for Astrophysics and Space Astronomy,
Department of Astrophysical and Planetary Sciences,
University of Colorado, 389 UCB,
Boulder, CO 80309-0389, USA}
\author{Jennie Paine}
\affil{Center for Astrophysics and Space Astronomy,
Department of Astrophysical and Planetary Sciences,
University of Colorado, 389 UCB,
Boulder, CO 80309-0389, USA}



\begin{abstract}
The canonical methods for gravitational wave detection are ground- and 
space-based laser interferometry, pulsar timing, and 
polarization of the cosmic microwave background.  But as has been suggested
by numerous investigators, astrometry offers an additional path to gravitational 
wave detection.  Gravitational waves deflect light rays of extragalactic objects,
creating apparent proper motions in a quadrupolar (and higher-order modes) pattern.  
Astrometry of extragalactic radio sources is sensitive to gravitational waves with frequencies between roughly $10^{-18}$ and 
$10^{-8}$ Hz ($H_0$ and 1/3 yr$^{-1}$), overlapping and bridging the pulsar timing and CMB polarization regimes.
We present a methodology for astrometric gravitational wave detection in the 
presence of large intrinsic uncorrelated proper motions (i.e., radio jets).
We obtain 95\% confidence limits on the stochastic gravitational wave background using 711 radio 
sources, $\Omega_{\rm GW} < 0.0064$, and using 508 radio sources combined with the 
first {\it Gaia} data release:  $\Omega_{\rm GW} < 0.011$.  These limits probe gravitational wave frequencies 
$6\times10^{-18}$~Hz $\lesssim f \lesssim 1\times10^{-9}$~Hz.  
Using a {\it WISE}-{\it Gaia} catalog of 567,721 AGN, we predict a limit expected from {\it Gaia} alone
of $\Omega_{\rm GW} < 0.0006$, which is significantly higher than was originally forecast.
Incidentally, we detect and report on 22 new examples of optical superluminal motion with redshifts
0.13--3.89.
\end{abstract}




\section{Introduction} \label{sec:intro}

A stochastic gravitational wave background deflects light from distant objects, producing an apparent 
proper motion \citep{braginsky1990}.  The angular deflections will be correlated across the sky with an amplitude of 
the order of the dimensionless strain of the gravitational waves, $h_{\rm rms}$ \citep{braginsky1990,kaiser1997};
one microarcsecond ($\mu$as) 
of deflection is equivalent to a dimensionless strain $h\sim5\times10^{-12}$.  
Observations spanning a time
interval $\Delta t = 1/f_{obs}$ will be sensitive to gravitational waves with frequencies $f < f_{obs}$, roughly down to 
the inverse of the light travel time to the observed objects, $f\sim 10^{-18}$--$10^{-17}$ Hz \citep[e.g.,][]{book11}.  

\citet{book11} show that the cosmological gravitational wave background energy 
density can be related to the correlated light deflections as 
\begin{equation}
  \Omega_{\rm GW} (f) \sim \langle\mu(f)^2\rangle/H_0^2,
\end{equation}
where $\langle\mu(f)^2\rangle$ is the variance in the proper motion at observed frequency $f$, 
and  $H_0$ is the Hubble constant.
The proper motion power spectrum, for quadrupolar and higher-order modes, can measure or constrain the 
gravitational wave background over 10 decades in frequency, $H_0 \lesssim f \lesssim 1$~yr$^{-1}$ 
($10^{-18} \lesssim f \lesssim 10^{-8}$ Hz).
The dominant signal is quadrupolar (Fig. \ref{fig:random_quad}), with smaller contributions from $\ell > 2$ modes.
In practice, observations constrain the energy density integrated over frequencies, $\int \Omega_{\rm GW}(f)\, d \ln f$, 
which will hereafter be labeled as $\Omega_{\rm GW}$.

Detecting or constraining $\sim$10$^{-18}$--$10^{-16}$ Hz primordial gravitational waves is a key goal of 
CMB polarization $B$-mode measurements \citep[e.g.,][]{kamionkowski1997,seljak1997,litebird2016}, but between the CMB
polarization measurements and pulsar timing, which is sensitive to frequencies $\sim$10$^{-9}$--$10^{-7}$ Hz 
\citep[e.g.,][]{nanograv2016}, are $\sim$7 orders of magnitude in frequency space.  Proper motion measurements can approach 
gravitational wave detection in a completely independent manner and bridge the frequency gap between
the pulsar timing and CMB polarization methods.  The frequency range $f\gtrsim10^{-15}$~Hz can also be probed using the 
CMB power spectrum because gravitational waves contribute to the radiation density of the universe and can mimic
a massless neutrino, modifying the effective number of neutrino species \citep[e.g.,][]{smith2006}.  

Previous observational work on astrometric detection of gravitational waves using active galactic nuclei (AGNs)
using radio interferometry include \citet{gwinn97} and \citet{titov11}.  They quote upper limits on the 
stochastic gravitational wave background --- expressed in terms of the critical cosmological energy density --- 
of $\Omega_{\rm GW} < 0.11\ h_{100}^{-2}$
for $f < 2\times10^{-9}$ Hz at 95\% confidence \citep{gwinn97} and $\Omega_{\rm GW} < 0.0042\ h_{100}^{-2}$
for $f < 10^{-9}$ Hz \citep{titov11}.  However, we cannot reproduce either of these limits based on their quoted quadrupolar fit 
parameters.

In this paper, we present detailed methods for astrometric measurement of the gravitational wave background 
including a maximum likelihood method for extracting correlated signals in vector fields with large significant 
outliers (the uncorrelated ``intrinsic'' apparent proper motion induced by relativistic jets; Section \ref{sec:methods}).  
We use these methods in Sections \ref{sec:data}--\ref{sec:results} to obtain new stochastic gravitational wave limits from a 
Very Long Baseline Array (VLBA)\footnote{The National Radio Astronomy 
Observatory is a facility of the National Science Foundation operated under cooperative agreement by Associated Universities, Inc.} 
astrometric catalog \citep{truebenbach2017} and from this catalog combined with the first {\it Gaia} data release \citep{gaia2016a,gaia2016b}.
We also use a {\it Gaia}-{\it WISE} catalog \citep{paine2018} to make predictions for the gravitational wave detection sensitivity of 
{\it Gaia} by the end of its mission (Section \ref{sec:predictions}).  

The only cosmological assumption used for the gravitational wave results is  $H_0 = 70$ km s$^{-1}$ Mpc$^{-1}$.  When expressed 
as an angular frequency, the Hubble constant becomes $H_0 = 15$~$\mu$as~yr$^{-1}$.  
For superluminal motion\footnote{See \citet{cohen1977}, \citet{blandford1977}.}
 calculations, we additionally assume a flat cosmology with $\Omega_M = 0.27$, and 
$\Omega_\Lambda = 0.73$.

\section{Methods}\label{sec:methods}

To characterize a vector field on a sphere, 
one can extend the usual spherical harmonic characterization of a scalar field on a sphere to 
vector spherical harmonics \citep[e.g.,][]{thorne1980}, defined as the gradient and curl of the scalar spherical harmonics, 
which resemble electric ($E$) and magnetic ($B$) fields \citep{mignard12}:
\begin{equation}
\vec{S}_{\ell m}(\alpha,\delta) = {1 \over \sqrt{\ell(\ell+1)}}\ \vec{\nabla} Y_{\ell m}(\alpha,\delta),
\end{equation}
and
\begin{equation}
\vec{T}_{\ell m}(\alpha,\delta) = {-1 \over \sqrt{\ell(\ell+1)}}\ \hat{n} \times \vec{\nabla} Y_{\ell m}(\alpha,\delta),
\end{equation}
where the $\vec{S}_{\ell m}$ is the ``spheroidal'' $E$-mode of degree $\ell$ and order $m$, 
$\vec{T}_{\ell m}$ is the ``toroidal'' $B$-mode, and $\hat{n}$ is the radial unit vector.  
$\vec{S}_{\ell m}$, $\vec{T}_{\ell m}$, and $\hat{n}$ are mutually orthogonal, by construction.
A general vector field $\vec{V}(\alpha,\delta)$ on the surface of a sphere can be expanded in terms of this 
vector spherical harmonic basis using complex coefficients $s_{\ell m}$ and $t_{\ell m}$:
\begin{equation}\label{eqn:vsh}
  \vec{V}(\alpha,\delta) = \sum_{\ell=1}^{\infty} \sum_{m=-\ell}^\ell
                   (s_{\ell m} \vec{S}_{\ell m}(\alpha,\delta) + t_{\ell m} \vec{T}_{\ell m}(\alpha,\delta) ).
\end{equation}
The first three spherical harmonic degrees relevant to this treatment ($E$- and $B$-mode dipole, quadrupole, and octopole)
are listed in Tables \ref{tab:Emultipoles} and \ref{tab:Bmultipoles} and explicitly as equations with coefficients in Appendix 
\ref{appendix:vsh} (but note that these equations describe a real-valued vector field and are a special case of
the general complex vector spherical harmonics described by Equation \ref{eqn:vsh}).
We follow the \citet{mignard12} prescriptions for calculating the power in any mode (the quadrature sum of 
coefficients, modulo factors of 2),
\begin{equation}\label{eqn:power}
  P_\ell = s_{\ell 0}^2 + t_{\ell 0}^2 +2 \sum_{m=1}^{\ell} \left( (s_{\ell m}^{Re})^2+ (s_{\ell m}^{Im})^2+ (t_{\ell m}^{Re})^2+ (t_{\ell m}^{Im})^2\right),
\end{equation}
 and use the $Z$-score to assess significance \citep[][Eqn.\ 85]{mignard12}.

The \citet{gwinn97} power (sum of squared ``moduli'' [amplitudes]) is equivalent to the \cite{mignard12} 
power prescription, despite slightly different definitions.  One can therefore use the quadrupole power ($\ell = 2$)
as described in Equation \ref{eqn:power} to obtain an estimate of the gravitational wave energy density:
\begin{equation}\label{eqn:Omega}
    \Omega_{\rm GW} = {6\over5}\,{1\over4\pi}\,{P_2\over H_\circ^2} 
                              = 0.00042\, {P_2\over (1\ \mu\rm{as\ yr}^{-1})^2}\,h_{70}^{-2}
\end{equation}
The factor of $6/5$ in this expression corrects for the $5/6$ contribution of the quadrupole to the 
total gravitational wave signal \citep{gwinn97,book11}.

\begin{figure}[t]
\epsscale{1.18}
\plotone{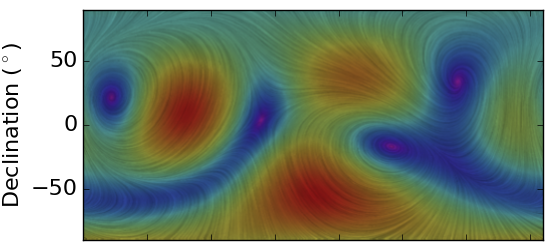}
\plotone{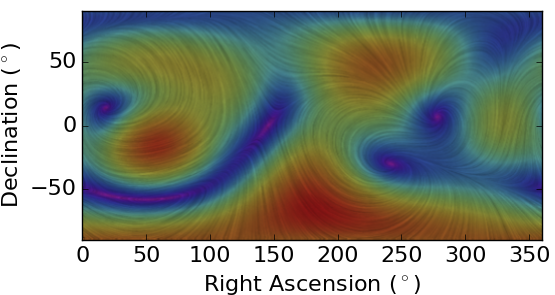}
\caption{\footnotesize
Randomly generated all-sky quadrupole (top) and quadrupole plus octopole (bottom) 
$E$- and $B$-mode stream plots in equatorial coordinates (see Section \ref{sec:methods}).
For the combined quadrupole and octopole plot, the weighting is $(5/6)^{1/2}$ and $(7/60)^{1/2}$, respectively,
which accounts for 95\% of the expected signal power \citep[the remainder is in higher multipole modes;][]{book11}. 
Streamlines indicate the vector field direction, and the colors indicate the vector amplitude, from violet (zero) to red (maximum).}\label{fig:random_quad}
\end{figure}

In general, a proper motion catalog that produces a limit on the quadrupole vector spherical harmonics 
can also provide a similar limit on the octopole (and higher orders), but the expected relative weighting on 
quadrupole power versus higher multipoles in a stochastic gravitational wave signal declines rapidly, as  
$\ell^{-4.9}$ \citep{book11}.  A quadrupole-only limit will typically be the most constraining, despite the additional
information contained in higher-order modes, so a ``bandpower'' approach such as that used in CMB 
signal detection will not be effective \citep[e.g.,][]{bond1998}.  We demonstrate this explicitly using data in Section \ref{sec:results}.
Figure \ref{fig:random_quad} compares an $\ell=2$ proper motion stream plot 
to a $(5/6)^{1/2}\ \vec{V}_2 + (7/60)^{1/2}\ \vec{V}_3$ vector field.  The quadrupole and octopole coefficients 
were randomly selected from normal distributions with the appropriate $\sqrt{2}$ scaling of the $m=0$ terms.  
The differences between the two cases are subtle because the octopole power is de-weighted by a factor of $\sim$7
compared to the quadrupole.

\section{Data Sources and Proper Motions}\label{sec:data}

We measure proper motions from astrometric time series using VLBA data only and VLBA data combined with 
a single {\it Gaia} epoch.  For both time series, 
we fit position versus time in R.A. and decl.\ separately using error-weighted linear least-squares
bootstrapped to incorporate the effect of outlier epochs, as described in \citet{truebenbach2017}.

\subsection{VLBA Catalog}\label{subsec:vlba}

The VLBA astrometric catalog is described and characterized in detail in \citet{truebenbach2017}.  In summary, 
the catalog contains 713 objects with mean astrometric uncertainties of 24 $\mu$as yr$^{-1}$.  These
were obtained from long-term astrometric monitoring programs as well as new observations.  Proper motions 
were measured from astrometric time series using a bootstrapped error-weighted least-squares fit for each 
object in each coordinate, substantially improving on previous proper motion measurements for most objects.  

\begin{figure}[t]
\epsscale{1.18}
\plotone{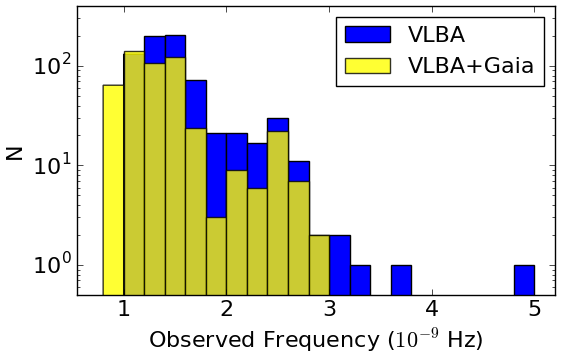}
\caption{\footnotesize
Distribution of observed frequencies obtained from astrometric time series of the VLBA and VLBA+{\it Gaia}
catalogs.  The upper bound on the gravitational wave frequency sensitivity of an object is the inverse of the 
time span used to measure its proper motion.  The gravitational wave frequencies probed by these
proper motions span the range  $6\times10^{-18}$~Hz~$\lesssim f \lesssim 1\times10^{-9}$~Hz 
(see Section \ref{subsec:vlba}).   
}\label{fig:frequencies}
\end{figure}

The time baselines spanned by the new and archival data cover the range 6.4--27.2 years 
($1.2\times10^{-9}$ Hz to $5.0\times10^{-9} $ Hz).  The median time 
series spans 22.2 years, which is equivalent to $f = 1.4\times10^{-9}$ Hz.  Figure \ref{fig:frequencies} shows
the distribution of observed frequencies for the catalog.  The majority (95\%) of objects sample the 
range $f_{\rm obs} = (1.0$--$2.5)\times10^{-9} $ Hz.  The lower bound on detectable frequencies is 
set by the distance of the catalog objects, which must be greater than the wavelength of the gravitational waves.  
The sensitivity of the sample to the longest wavelength gravitational waves 
will therefore be a function of frequency because the sample size decreases with distance.   
The median redshift is 1.10, and the upper and lower quartile 
divisions are 0.594 and 1.64.  We conservatively set the lower
bound on frequency using the first redshift quartile, $z_{25\%} = 0.594$, 
where 75\% of the sample can still be used to detect gravitational 
waves.  Using our assumed cosmology, the light travel time is 5.74 Gyr, which corresponds to 
$6\times10^{-18}$ Hz.  
The astrometry is therefore sensitive to gravitational waves
with $6\times10^{-18}$~Hz~$\lesssim f \lesssim 1\times10^{-9}$~Hz.

While this catalog shows very low proper motion errors, the proper motions themselves 
can be substantial and significant due to relativistic radio jet motion (see Section \ref{subsec:vlba-gaia}).  
This ``intrinsic'' proper motion is uncorrelated between objects, but introduces 
special challenges to detecting small-amplitude correlated global proper motions.  Section \ref{sec:signals}
presents a solution to this uncorrelated large-amplitude significant-signal contamination problem.  

\subsection{VLBA+Gaia Catalog}\label{subsec:vlba-gaia}

The first {\it Gaia} data release (DR1) catalog \citep{gaia2016a} contains a single-epoch (2015.0) position 
for 2191 AGN in the International Celestial Reference Frame (ICRF2) catalog \citep{mignard2016}.  
Five-hundred seventy-seven (577) of these are VLBA sources in the \citet{truebenbach2017} catalog, and we use them to measure
proper motions from the VLBA-{\it Gaia} time series.  Median uncertainties in the {\it Gaia} astrometry
of these objects are 518 $\mu$as and 459 $\mu$as in R.A. and decl., respectively.

To create a VLBA+{\it Gaia} proper motion catalog, we perform a 500-iteration bootstrap error-weighted 
least-squares fit to the VLBA time series as described by \citet{truebenbach2017}, but we include the 
{\it Gaia} point in every fit rather than allowing it to fall into the bootstrap selection pool.  This causes the {\it Gaia}
point to act as a loose astrometric anchor, to within its uncertainty.  When there are many radio epochs in a time 
series, the {\it Gaia} point will still have a minor impact on the best-fit proper motion.

\begin{figure*}
\epsscale{1.15}
\plotone{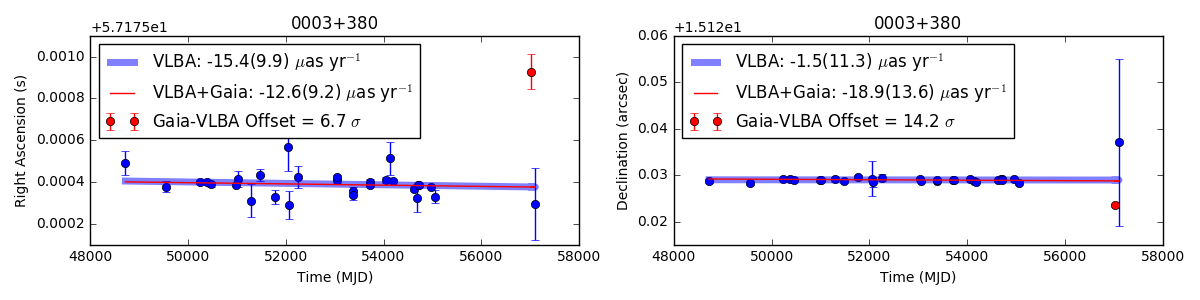}
\plotone{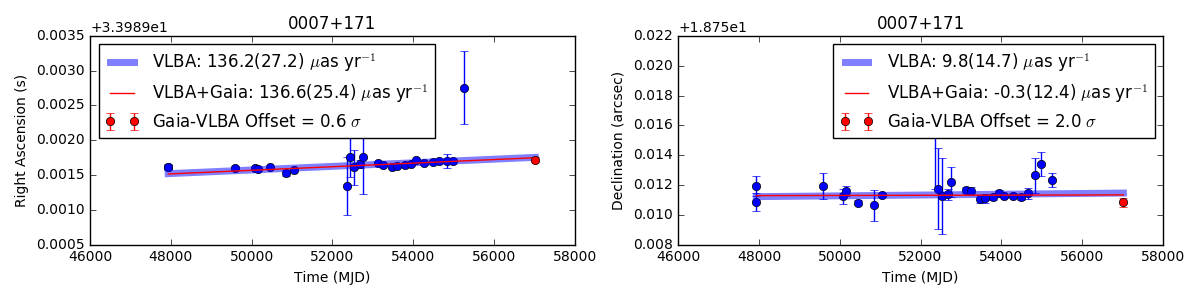}
\plotone{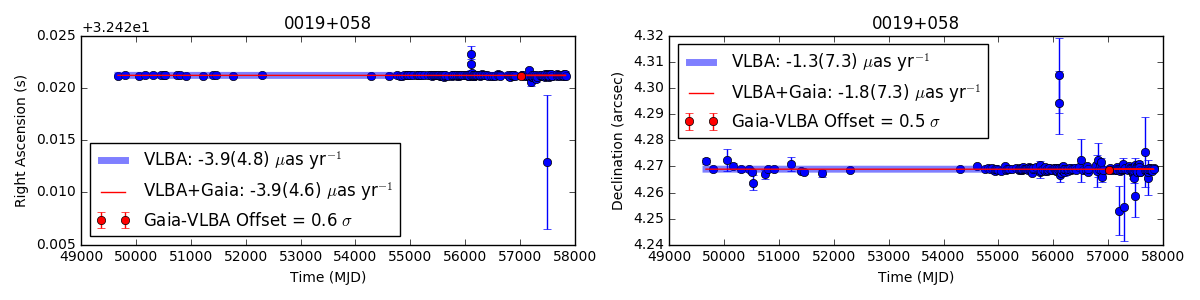}
\caption{\footnotesize
Example time series astrometric fits of VLBA only (blue) and VLBA plus the 2015.0 
{\it Gaia} epoch (red).
The columns depict the R.A. (left) and the decl.\ (right), and the 
proper motions and their errors are indicated in the inset boxes.
Top panels:  0003+380 shows significant {\it Gaia}-VLBA inconsistency, indicating physical offsets between 
the radio and optical emission regions.
Middle panels:  0007+171 shows highly significant (and superluminal) motion where the optical and radio 
emission regions are consistent.  This is also an example of the {\it Gaia} epoch substantially extending the time 
series.  
Bottom panels:  0019+058 shows agreement between the VLBA and {\it Gaia} astrometry where no proper motion is 
detected.  This case represents the majority (77\%) of the astrometric sample.  
}\label{fig:timeseries}
\end{figure*}

We assess the {\it Gaia} offset from the VLBA-only proper motion prediction for the {\it Gaia} epoch 
strictly based on the uncertainty in the {\it Gaia} measurement, which is typically larger than the 
prediction uncertainty of the time series fit.  
The proper motions obtained from the time series that include a single {\it Gaia} epoch are typically not 
significantly altered from the VLBA-only results, but there are some notable exceptions.  
It is remarkable that 88\% of the objects' proper motions show consistency between the radio trend 
and the {\it Gaia} position.  In most of these cases (87\% of the consistent subset), 
the measured proper motion, given its uncertainty, is
consistent with zero, which implies that the optical and radio positions coincide with no motion to within the 
measurement uncertainty.  

In 13\% of the sample, the {\it Gaia} epoch extends the time series beyond the VLBA epochs, 
agrees with the VLBA proper motion, 
and generally improves the proper motion solution (see 0007+171 in Figure \ref{fig:timeseries}).  
We assume in these cases that the optical and radio 
centroids are coincident (as is the case with most objects in the sample).

In 12\% of the sample, the {\it Gaia} epoch is significantly ($> 3 \sigma$) offset from the VLBA proper motion 
fit in one or both coordinates, indicating that the optical and radio centroids do not coincide.  
Figure \ref{fig:timeseries} shows an example, 0003+380, where the {\it Gaia} astrometry is significantly offset in both coordinates.
We cannot yet say whether the 
radio and optical proper motions differ because there is only one optical astrometric epoch.   There are many possible 
reasons for radio-optical offsets, including optically faint jet emission, optical light from the host galaxy, dust obscuration 
of the AGN, and offset radio and optical emission regions within jets \citep[e.g.,][]{kovalev2016}.  
Given these scenarios, it is surprising that 88\% of the sample 
does show good radio-optical coincidence at the sub-mas level \citep[but see][]{petrov2017}.  

In 9\% of the subsample with good VLBA+{\it Gaia} agreement (8\% of the total sample), the proper motion is significant ($> 5 \sigma$ in at least one coordinate)
and intrinsic to the object (not cosmological, caused by gravitational waves, or observer-induced).  
For example, 0007+171 shows a R.A. proper motion of 136.6(25.4) $\mu$as yr$^{-1}$ (Figure \ref{fig:timeseries} and Tables \ref{tab:superluminal} and \ref{tab:pms}).
At a redshift of $z=1.601$ \citep{wills1976},
this proper motion coincides with apparent superluminal motion of $10.1(1.9)c$ in the object's rest frame.

Apparent velocities are calculated in the source rest frame using the proper motion distance $D_M$, which is equal to the line-of-sight
comoving distance in a flat cosmology and related to the angular diameter distance $D_A$ as $D_M = D_A (1+z)$ \citep{hogg1999}.  
The redshift scale
factor translates the observer-frame time interval used to calculate apparent velocity into the object rest-frame time interval.  We therefore
calculate the apparent transverse velocity from proper motion via
\begin{equation}
  \vec{v} = \vec{\mu}\ D_M
\end{equation}
where $D_M$ may be expressed in distance per radian or most often kpc arcsec$^{-1}$. 

In total, there are 22 objects that show superluminal optical and radio motion, up to 10.1 $c$.  
Table \ref{tab:superluminal} lists the properties of these objects.
We interpret these observations to indicate that the AGN radio jets also show significant detectable 
optical emission, even at 
substantial redshifts.  It is noteworthy that optical superluminal motion has only been observed in the local universe
in a small number of objects including M87 and 3C264 \citep[e.g.,][]{biretta1999,meyer2015}. 

Table \ref{tab:pms} lists the VLBA-only and the VLBA+{\it Gaia} proper motions as well as the {\it Gaia} offsets
from the VLBA-only fits.  
For signal extraction from this catalog, we exclude objects with 
significant (3$\sigma$) {\it Gaia} offsets in either coordinate direction.  
After culling, 508 objects remain in this sample, and these are used 
in the vector spherical harmonic fits  (Sections \ref{sec:signals} and \ref{sec:results}).

The time baselines spanned by the VLBA+{\it Gaia} proper motion catalog are somewhat longer than 
the VLBA-only catalog (see Figure \ref{fig:frequencies}).   Time series range from 10.6 to 37.6 years
($8.4\times10^{-10}$ Hz to $3.0\times10^{-9} $ Hz), and the median time 
series spans 24.9 years, which is equivalent to $f = 1.3\times10^{-9}$ Hz.  The majority (95\%) of objects sample the 
range $f_{\rm obs} = 0.8$--$2.5\times10^{-9} $ Hz.  
The median redshift is 1.23, and the upper and lower quartile 
divisions are 0.73 and 1.80.  The first redshift quartile, $z_{25\%} = 0.73$, with light travel time 6.55 Gyr,
sets the lower bound on frequency of $5\times10^{-18}$ Hz.  
The $\ell \geq 2$ correlated proper motions used to constrain the 
stochastic gravitational wave background are therefore sensitive to waves with frequencies
$5\times10^{-18}$~Hz~$\lesssim f \lesssim 0.8\times10^{-9}$~Hz.

\subsection{Gaia Catalog}\label{subsec: gaiacat}

The advantages of extragalactic {\it Gaia} proper motions over radio interferometric proper motions lie in the factor of 
$\sim$1000 increase in number of optical sources over radio 
and the (generally) lower intrinsic optical proper motions.  These advantages may overcome the 
less precise astrometry and shorter time baseline of {\it Gaia} compared to geodetic VLBI monitoring (in the short term).

The {\it Gaia}  DR1 catalog contains a single-epoch position for AGN, but the 
expected end-of-mission proper motion uncertainties can be used to predict the
sensitivity of the final {\it Gaia} catalog to gravitational waves. 
To first order, the vector proper motion error of each object depends on its ecliptic angle and optical 
{\it G}-band magnitude\footnote{\url{https://www.cosmos.esa.int/web/gaia/science-performance}}.  
We use the {\tt pyGaia}\footnote{\url{https://pypi.python.org/pypi/PyGaia}} 
package to predict the proper motion errors for each object in the 567,721 AGN \citet{paine2018} {\it WISE}-{\it Gaia} 
sample.  \citet{paine2018} present the expected uncertainties, the sky distribution, and the 
potential systematics of the sample.  We use this catalog in Section \ref{sec:predictions} 
to predict the expected {\it Gaia} end-of-mission sensitivity to the stochastic gravitational wave 
background.

\section{Signal Extraction}\label{sec:signals}

The challenge to vector spherical harmonic fitting posed by radio sources is their 
often significant large apparent intrinsic proper motions induced by relativistic jets.  
These intrinsic proper motions are uncorrelated between objects but can dominate 
an error-weighted least-squares fit of the correlated proper motions.  
Investigators measuring the secular aberration drift dipole have therefore heavily 
censored their samples in order to maximize the 
signal of interest \citep[e.g.,][]{titov11,titov13}, but this requires {\it a priori} knowledge of
the expected signal.  A different approach can be bootstrap resampling, which was successfully implemented 
by \citet{truebenbach2017} to extract the secular aberration drift dipole induced
by the barycenter acceleration about the Galactic Center with minimal data clipping.  

Here, we implement a maximum likelihood MCMC ``permissive fit'' method that
allows for highly significant large-departure data points by assuming that 
the mismatch between model and data will in some cases be bounded from below
by the measured uncertainty \citep[][p.\ 168]{sivia2006}.
This method, rather than minimizing an error-weighted data-model residual 
$R_i = (D_i - \rm{Model})/\sigma_i$ for each 
data point $D_i$ with uncertainty $\sigma_i$,
maximizes the logarithm of the posterior probability density function
\begin{equation}
  L = {\rm constant} + \sum_{i=1}^{N}\  \ln \left( 1-e^{-R_i^2/2} \over R_i^2 \right).
\end{equation}
In this work, the data are the positions and proper motions of extragalactic 
objects, the model is a linear combination of vector spherical harmonics evaluated
at each object position, and the 
uncertainties are the proper motion errors (uncertainties in the positions of objects
have no impact on low-$\ell$ signals).  
To assess the model fits and uncertainties, we employ an MCMC technique using 
{\tt lmfit} \citep{newville2014} to obtain the maximum likelihood and 
confidence intervals for each fit parameter directly from the resulting distribution of outcomes.  

For the vector spherical harmonic fits, all coefficients and uncertainties are maximum likelihood estimates.
The power in a given mode is calculated from Equation \ref{eqn:power}, and its significance is estimated 
using a $Z$-score following \citet{mignard12}, Eqn.\ 85.

\subsection{The Secular Aberration Drift Dipole}

We start with a fit of the $E$- and $B$-mode dipole signals in both catalogs, which must be removed from vector
fields before attempting to measure higher-order modes.  While the vector spherical harmonics are orthonormal in 
principle, there can be correlation between degrees and orders when fitting the harmonics to 
discrete sparsely sampled nonuniform noisy data, so ``nuisance'' signals must be subtracted.
We simultaneously fit for both $E$- and $B$-mode dipoles (aberration drift and 
rotation, respectively) to capture any residual signature of a non-inertial frame
(or other cosmic rotation) as well as any correlations between the two modes
(even if the $B$-mode dipole is nonsignificant, there can still be crosstalk with the $E$-mode dipole).  
Following \citet{mignard12}, the dipole equations
are listed in Tables \ref{tab:Emultipoles} and \ref{tab:Bmultipoles} and explicitly as equations with coefficients in Appendix 
\ref{appendix:vsh}.  

\begin{figure}[t]
\epsscale{1.23}
\plotone{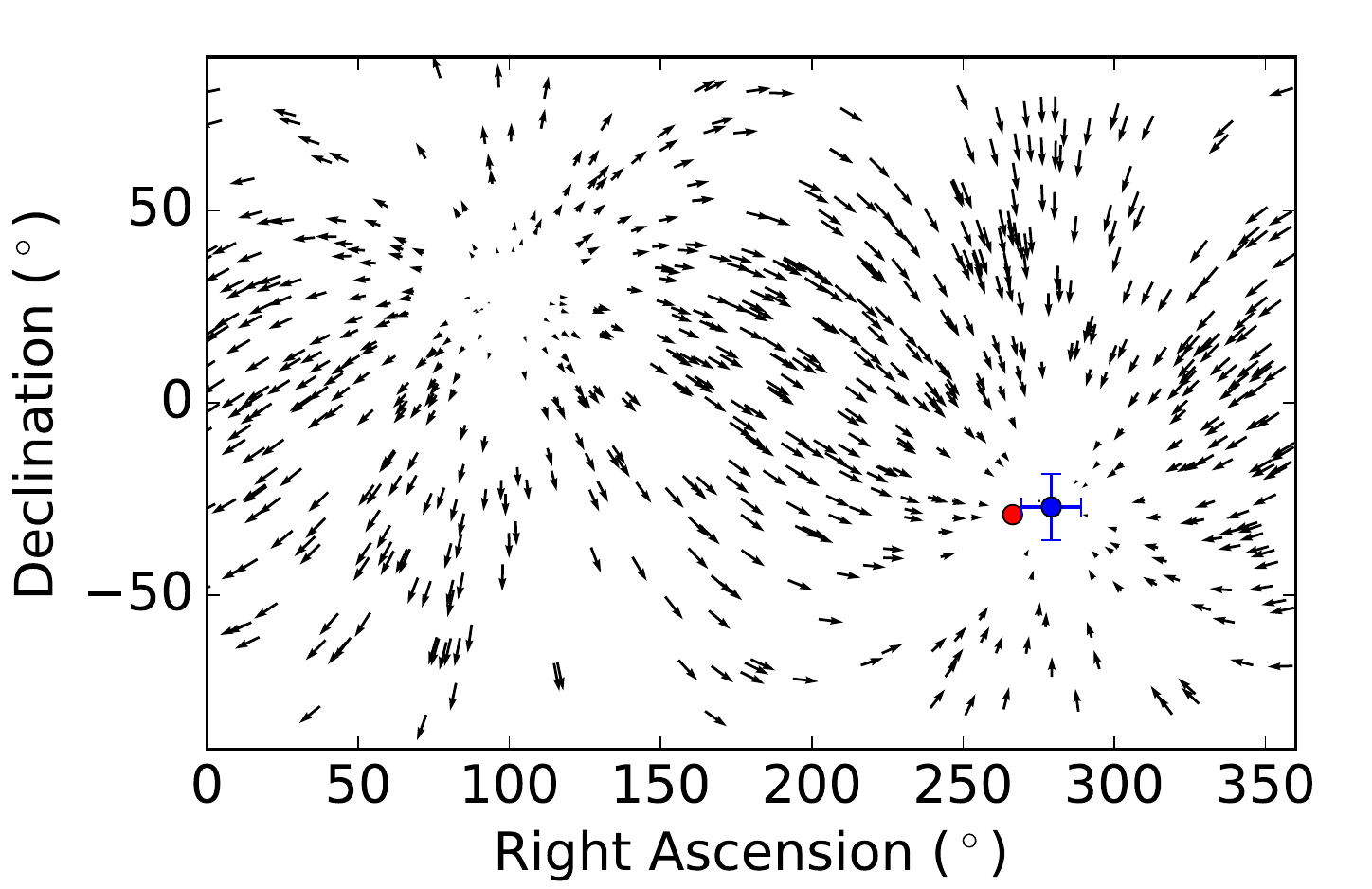}
\caption{\footnotesize
Maximum likelihood secular aberration drift ($E$-mode dipole) model fit to the VLBA sample
plotted in equatorial coordinates.  
The fit parameters and uncertainties are listed in Table \ref{tab:dipole}. 
The red circle indicates the Galactic Center, and the point with the error bars shows the dipole apex
obtained from the dipole (simultaneous $E$- and $B$-mode) fit.  
}\label{fig:dipole}
\end{figure}

Table \ref{tab:dipole} lists the dipole fit coefficients for the VLBA and VLBA+{\it Gaia} samples.  
We significantly detect the secular aberration drift with 5.5$\sigma$ and 5.1$\sigma$ significance, 
respectively.  The $E$-mode dipole apex lies at $279\fdg2(9\fdg8)$, $-27\fdg0(8\fdg7)$ and
$274\fdg0(9\fdg3)$, $-18\fdg0(9\fdg2)$, and 
is consistent with the Galactic Center ($266\fdg4$, $-29\fdg0$) in each case (Figure \ref{fig:dipole}). The dipole  
amplitude is 1.70(0.26) and 1.70(0.29) $\mu$as yr$^{-1}$, which is substantially smaller than expected. 
For example, \citet{titov13} find 6.4(1.1) $\mu$as yr$^{-1}$, and \citet{xu2013} obtain 5.8(0.3) $\mu$as yr$^{-1}$. These are
consistent with the expectation of 5.5(0.2) $\mu$as yr$^{-1}$ based on barycenter orbital parameters obtained from 
Galactic maser parallaxes and proper motions and the Sgr A* reflex motion \citep{reid14}.
The amplitude of the dipole measured here is likely suppressed by the no-net-rotation
constraint imposed by the global fitting used to produce the ICRF catalog (see \citet{truebenbach2017} 
for a detailed discussion). 
For the purposes of detecting higher multipole modes in the proper motion data, we simply need to 
measure whatever dipole is present and subtract the dipole fit from each proper motion catalog before fitting higher multipoles to 
extract (or constrain) the gravitational wave signal.  

No rotation is detected:  the $B$-mode dipole fits are not significant (1.2$\sigma$ or less), with square root power 
of 1.89(0.74) $\mu$as yr$^{-1}$ for the VLBA catalog and 1.29(0.79)  $\mu$as yr$^{-1}$ for VLBA+{\it Gaia}.  
These correspond to angular rotation rates of 0.65(0.26) and 0.45(0.27)  $\mu$as yr$^{-1}$, respectively.

\begin{deluxetable}{lrr}
\tabletypesize{\scriptsize} 
\tablecaption{Dipole Fits \label{tab:dipole}}
\tablewidth{0pt} 
\tablehead{
\colhead{} &
\colhead{VLBA} &
\colhead{VLBA+{\it Gaia} }\\
\hline
\colhead{Quantity} &
\colhead{Amplitude} &
\colhead{Amplitude} \\
\colhead{} & \colhead{($\mu$as yr$^{-1}$)} & 
\colhead{($\mu$as yr$^{-1}$)}}
\startdata 
\multicolumn{3}{l}{$E$-Mode Dipole (Aberration Drift)}\\
$s_{10}$       & $-2.24(0.74)$ & $-1.52(0.78)$ \\   
$s_{11}^{Re}$ & $-0.50(0.53)$ & $-0.21(0.53)$ \\
$s_{11}^{Im}$ & $-3.07(0.54)$ & $-3.29(0.61)$ \\
\boldmath{$\sqrt{P_1^s}$} & \boldmath{$4.93(0.76)$} & \boldmath{$4.91(0.85)$}\\
Z-score\tablenotemark{a} & 5.5 & 5.1\\
\hline
\noalign{\vskip 1mm} 
\multicolumn{3}{l}{$B$-Mode Dipole (Rotation)}\\
$t_{10}$       & $-0.72(0.62)$ & $-0.51(0.68)$ \\
$t_{11}^{Re}$ & $+1.17(0.52)$ & $0.76(0.55)$\\
$t_{11}^{Im}$ & $-0.40(0.66)$ & $-0.36(0.68)$\\
\boldmath{$\sqrt{P_1^t}$} & \boldmath{$1.89(0.74)$} & \boldmath{$1.29(0.79)$}\\
Z-score\tablenotemark{a} & 1.2 & $0.1$\\
\enddata 
\tablenotetext{a}{This statistic is unitless.}
\tablecomments{Fits are simultaneously made to electric and magnetic dipole vector fields.}
\end{deluxetable}

\subsection{Constraints on Gravitational Waves}\label{subsec:GWs}

For each catalog, we simultaneously fit $E$- and $B$-mode quadrupolar vector spherical harmonics for the best 
constraint on the stochastic gravitational wave background.  We also simultaneously
fit $E$- and $B$-mode quadrupole and octopole coefficients, but this fit is less constraining than 
quadrupole alone, as described in Section \ref{sec:methods} and demonstrated 
explicitly in Section \ref{sec:results}.  
Measuring the quadrupole and octopole powers separately is a way to test the isotropy of the 
background, which is assumed when combining the two modes to obtain a limit on $\Omega_{\rm GW}$.
The quadrupole and octopole vector spherical harmonics 
are listed in Tables \ref{tab:Emultipoles} and \ref{tab:Bmultipoles} and explicitly as equations 
with coefficients in Appendix \ref{appendix:vsh}, Equations \ref{eqn:Equad}--\ref{eqn:Boct}.

It is incorrect to obtain a limit on $\Omega_{\rm GW}$ from 
the parameters of a nonsignificant quadrupole fit to a vector field.  When no significant signal is detected, 
we follow the method described by \citet{gwinn97}:  assuming independent Gaussian errors on the 
fit coefficients as determined using the 
above methods, we resample the fit components and recalculate the quadrupole power 10000 times.  
We set the upper limit to be 95th percentile of the quadrupole power distribution.  This method gives a 
less-constraining result than has been quoted in previous work, such as \citet{titov11}.

A stochastic gravitational wave background (or a reliable limit) should show equal power in the $E$- and $B$-modes
\citep{book11}.  One can therefore identify spurious, nongravitational wave signals by comparing the 
power in the two modes.

\section{Results}\label{sec:results}

\subsection{VLBA}

We fit the vector spherical harmonic quadrupole to 
the \citet{truebenbach2017} VLBA proper motion catalog with minimal 
restrictions on the fit sample.  We omit two objects with proper motion amplitudes 
greater than 1 milliarcsec yr$^{-1}$, leaving 711 objects in the catalog.
After $E$-mode dipole subtraction, a simultaneous $E$- and $B$-mode quadrupole fit produced no 
significant signal, with a total quadrupole power of $\sqrt{P_2} = 1.83(0.72)$ $\mu$as yr$^{-1}$.  
Table \ref{tab:quad} shows the fit components, mutually consistent (and nonsignificant) $E$- and $B$-mode 
powers ($\sqrt{P_2^s} = 1.46(0.69)$ $\mu$as yr$^{-1}$ and $\sqrt{P_2^t} = 1.11(0.77)$ $\mu$as yr$^{-1}$, 
respectively), and the $Z$-score of the fit.  Using the resampling method described above 
(Section \ref{subsec:GWs}), we obtain a 95\% confidence limit on the stochastic gravitational 
wave energy density of $\Omega_{\rm GW} < 0.0064$ (see Equation \ref{eqn:Omega}).  Direct conversion of 
the nonsignificant quadrupolar power to energy density gives a smaller (nonsignificant) value:  
$\Omega_{\rm GW} = 0.0014(0.0011)$.  This is roughly equivalent to a dimensionless gravitational wave
strain amplitude of $h\simeq10^{-10}$ for $f\simeq10^{-9}$ Hz \citep[$h\sim (H_0/f) \sqrt{\Omega_{\rm GW}}$;][]{book11}.
The maximum proper motion amplitude in the quadrupole is 1.0~$\mu$as~yr$^{-1}$, which is equivalent to 
$h\simeq10^{-10}$ for $f\simeq10^{-9}$ Hz ($h\sim \mu/f$).

A simultaneous quadrupole and octopole fit in both $E$- and $B$-modes produces a weaker
constraint on the stochastic gravitational wave background, as expected (see Section \ref{sec:methods}):
$\sqrt{P_2} = 3.53(1.09)$ $\mu$as yr$^{-1}$, $\sqrt{P_3} = 4.98(0.92)$ $\mu$as yr$^{-1}$, and 
$\Omega_{\rm GW} < 0.032$ (95\% confidence limit).  
To calculate the above limit on $\Omega_{\rm GW}$ from the nonsignificant quadrupole and octopole
fits, we substitute a weighted power into Equation \ref{eqn:Omega}, $1.05(P_2+P_3)$ in the 
place of $(6/5)P_2$, 
and resample the fit coefficients to find a 95\% confidence limit.  Table \ref{tab:quadoct}
lists the coefficients, mode powers, and $Z$-scores for this fit.

\begin{deluxetable}{lrr}
\tabletypesize{\scriptsize} 
\tablecaption{Quadrupole Fits \label{tab:quad}}
\tablewidth{0pt} 
\tablehead{
\colhead{} &
\colhead{VLBA} &
\colhead{VLBA+{\it Gaia}}\\
\hline
\colhead{Quantity} &
\colhead{Amplitude} &
\colhead{Amplitude} \\
\colhead{} & \colhead{($\mu$as yr$^{-1}$)} & 
\colhead{($\mu$as yr$^{-1}$)}}
\startdata 
\multicolumn{3}{l}{Quadrupole}\\
$s_{20}$       & $0.83(0.72)$ & $1.72(0.78)$ \\
$s_{21}^{Re}$ & $0.80(0.47)$ & $1.16(0.52)$ \\
$s_{21}^{Im}$ &$-0.22(0.48)$ & $-0.49(0.51)$ \\
$s_{22}^{Re}$ &$-0.17(0.58)$ & $-0.04(0.59)$ \\
$s_{22}^{Im}$ & $0.06(0.44)$ & $0.74(0.46)$ \\
\boldmath{$\sqrt{P_2^s}$} & \boldmath{$1.46(0.69)$} & \boldmath{$2.70(0.74)$} \\
$t_{20}$       &$-0.48(0.66)$ & $-1.17(0.70)$ \\
$t_{21}^{Re}$ & $0.01(0.50)$ & $-0.68(0.58)$ \\
$t_{21}^{Im}$ &$-0.16(0.43)$ & $-0.41(0.47)$ \\
$t_{22}^{Re}$ & $0.43(0.53)$ & $0.25(0.56)$ \\
$t_{22}^{Im}$ & $0.54(0.60)$ & $0.81(0.60)$ \\
\boldmath{$\sqrt{P_2^t}$} & \boldmath{$1.11(0.77)$} & \boldmath{$2.01(0.78)$}\\
\boldmath{$\sqrt{P_2}$} & \boldmath{$1.83(0.72)$} & \boldmath{$3.36(0.75)$}\\ 
Z-score\tablenotemark{a} & $-$0.7 & 1.9\\
\enddata 
\tablenotetext{a}{This statistic is unitless.}
\tablecomments{Fits are simultaneously made to electric and magnetic quadrupole vector fields.  Parenthetical quantities
indicate 1$\sigma$ uncertainties.}
\end{deluxetable}

\begin{deluxetable}{lrr}
\tabletypesize{\scriptsize} 
\tablecaption{Quadrupole and Octopole Fits \label{tab:quadoct}}
\tablewidth{0pt} 
\tablehead{
\colhead{} &
\colhead{VLBA} &
\colhead{VLBA+{\it Gaia}}\\
\hline
\colhead{Quantity} &
\colhead{Amplitude} &
\colhead{Amplitude} \\
\colhead{} & \colhead{($\mu$as yr$^{-1}$)} & 
\colhead{($\mu$as yr$^{-1}$)}}
\startdata 
\multicolumn{3}{l}{Quadrupole}\\
$s_{20}$       & $0.55(0.98)$ & $0.58(0.70)$\\
$s_{21}^{Re}$ & $0.88(0.62)$ & $0.29(0.71)$ \\
$s_{21}^{Im}$ &$-0.30(0.52)$ & $-0.99(0.61)$ \\
$s_{22}^{Re}$ &$-0.28(0.69)$ & $-0.42(0.76)$ \\
$s_{22}^{Im}$ & $0.04(0.30)$ & $0.05(0.54)$ \\
$t_{20}$       &$-0.27(0.20)$ & $-0.18(0.53)$ \\
$t_{21}^{Re}$ & $0.78(0.63)$ & $-0.10(0.57)$ \\
$t_{21}^{Im}$ & $0.13(0.36)$ & $-1.01(0.48)$ \\
$t_{22}^{Re}$ &$-0.50(0.37)$ & $0.02(0.01)$ \\
$t_{22}^{Im}$ & $2.05(0.84)$ & $1.66(0.83)$ \\
\boldmath{$\sqrt{P_2}$} & \boldmath{$3.53(1.09)$} & \boldmath{$3.23(1.01)$}\\
Z-score\tablenotemark{a} & 1.9 & 1.2\\
\hline
\noalign{\vskip 1mm} 
\multicolumn{3}{l}{Octopole}\\
$s_{30}$       &$-0.75(0.87)$ & $-0.34(1.02)$ \\
$s_{31}^{Re}$ &$-0.35(0.18)$ & $0.42(0.53)$ \\
$s_{31}^{Im}$ & $0.00(0.24)$ & $0.82(0.73)$ \\
$s_{32}^{Re}$ & $1.57(0.66)$ & $1.10(0.66)$ \\
$s_{32}^{Im}$ & $0.59(0.61)$ & $0.61(0.51)$ \\
$s_{33}^{Re}$ & $0.60(0.53)$ & $-0.37(0.52)$ \\
$s_{33}^{Im}$ & $0.85(0.58)$ & $0.39(0.58)$ \\
$t_{30}$       &$-0.10(0.81)$ & $-0.41(0.96)$ \\
$t_{31}^{Re}$ &$-1.43(0.58)$ & $-2.68(0.62)$ \\
$t_{31}^{Im}$ & $0.78(0.53)$ & $0.94(0.57)$ \\
$t_{32}^{Re}$ & $0.90(0.59)$ & $0.86(0.60)$ \\
$t_{32}^{Im}$ &$-0.77(0.49)$ & $-0.04(0.73)$ \\
$t_{33}^{Re}$ &$-2.01(0.75)$ & $-0.73(0.84)$ \\
$t_{33}^{Im}$ &$-0.12(0.61)$ & $-0.90(0.57)$ \\
\boldmath{$\sqrt{P_3}$} & \boldmath{$4.98(0.92)$} & \boldmath{$5.10(0.89)$}\\
Z-score\tablenotemark{a} & 3.4 & 2.6\\
\enddata 
\tablenotetext{a}{This statistic is unitless.}
\tablecomments{Fits are simultaneously made to electric and magnetic quadrupole and octopole vector fields.  
Parenthetical quantities indicate 1$\sigma$ uncertainties.}
\end{deluxetable}

\subsection{VLBA+Gaia}

We subtract the $E$-mode dipole and then fit the vector spherical harmonic quadrupole to 
the 508-object VLBA+{\it Gaia} proper motion catalog with no proper motion restrictions.
The simultaneous $E$- and $B$-mode quadrupole fit is not significant; the
total quadrupole power is $\sqrt{P_2} = 3.36(0.75)$ $\mu$as yr$^{-1}$, and the $E$- and $B$-mode
powers are consistent: $\sqrt{P_2^s} = 2.70(0.74)$ $\mu$as yr$^{-1}$ and 
$\sqrt{P_2^t} = 2.01(0.78)$ $\mu$as yr$^{-1}$, respectively.  
Table \ref{tab:quad} lists the fit components and the $Z$-score of the fit.  
The 95\% confidence limit on the stochastic gravitational 
wave energy density is $\Omega_{\rm GW} < 0.011$ (see Equation \ref{eqn:Omega}).

The simultaneous quadrupole and octopole fit in both $E$- and $B$-modes yields 
$\sqrt{P_2} = 3.23(1.01)$ $\mu$as yr$^{-1}$, $\sqrt{P_3} = 5.10(0.89)$ $\mu$as yr$^{-1}$, and 
$\Omega_{\rm GW} < 0.028$ (95\% confidence limit).  This limit on the stochastic gravitational
wave background is less constraining than the quadrupole-only fit, as expected, but is 
slightly more constraining than the VLBA-only quadrupole plus octopole fit.
Table \ref{tab:quadoct} lists the coefficients, mode powers, and $Z$-scores for this fit.

\section{{\it Gaia} Predictions}\label{sec:predictions}

The rough expectation for a stochastic gravitational wave 
background limit obtained from {\it Gaia} presented by \citet{book11} 
was $\Omega_{\rm GW} \lesssim 10^{-6}$ for $f\lesssim 10^{-8}$ Hz, but this was under the assumption of 
a proper motion uncertainty of $\sigma_\mu=10$~$\mu$as~yr$^{-1}$ per source and using
$10^6$ objects.  The revised end-of-mission error budgets, which have the largest impact on faint sources, have 
$\sigma_\mu\sim200$~$\mu$as~yr$^{-1}$ per quasar, and $\sim5\times10^5$ objects \citep{paine2018}, 
so we expect $\Omega_{\rm GW} \lesssim 10^{-3}$.  

To confirm this expectation, we use the expected end-of-mission proper motion errors on the 
extragalactic {\it Gaia}-{\it WISE} catalog compiled by \citet{paine2018} and 
described in Section \ref{subsec: gaiacat}.  We randomly sample vector proper 
motions from within the predicted error budget for each object assuming Gaussian errors.  
Provided that the barycenter acceleration about the 
Galactic Center can be removed from the data, the resulting proper motions should be uncorrelated 
and represent a no-signal noisy dataset that can be used to predict the best possible limit on a gravitational 
wave signal.  Unlike the {\it Gaia} Universe model snapshot (GUMS) sample \citep{robin2012}, this catalog
represents real objects detected in {\it Gaia} that will likely be employed when final proper motions are measured.  

After randomly sampling from within the proper motion error budgets for each object, we  
performed an error-weighted least-squares fit of 500 randomly generated
$E$- and $B$-mode quadrupolar gravitational wave signals with the quadrupolar power in the range 
$0.5$ $\mu$as yr$^{-1}\leq \sqrt{P_2} \leq 5$ $\mu$as yr$^{-1}$ in 25 logarithmic steps with 20 trials each.
For each trial, we add the input proper motion quadrupolar signal to the catalog proper motions and 
then fit simultaneous $E$- and $B$-mode quadrupole vector spherical harmonics (listed in 
Appendix \ref{appendix:vsh}) to obtain a fit power.  For each input power, 
we calculate the mean and standard deviation of the best-fit power to assess the offset and scatter of the fit
compared to the input.  

\begin{figure*}
\plotone{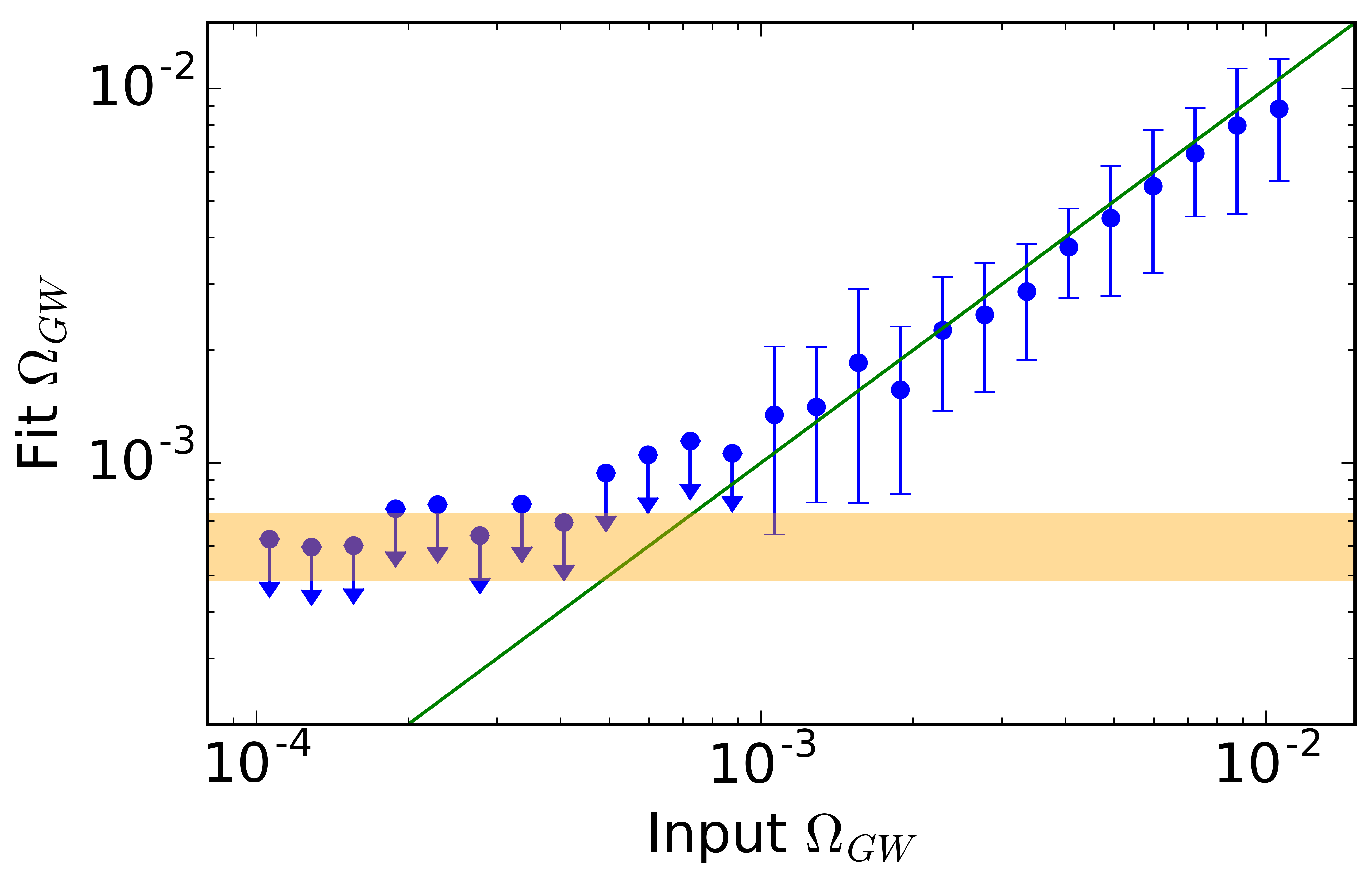}
\caption{Fit vs. input values for the stochastic gravitational wave background energy density, 
$\Omega_{\rm GW}$, expected from {\it Gaia} proper motions.  
Points and their error bars indicate the mean and standard deviation of 
20 recovery trials per random injected signal.  Arrows indicate nonsignificant fits.  
The green line indicates the one-to-one perfect signal recovery locus, and the 
orange bar shows the $\Omega_{\rm GW} = 6\times10^{-4}$  noise floor imposed by the end-of-mission {\it Gaia}
proper motion sensitivity to AGN in the {\it Gaia}-{\it WISE} catalog.}  \label{fig:GW_recovery}
\end{figure*}

Figure \ref{fig:GW_recovery} shows the results of these fit trials, recast in terms of 
$\Omega_{\rm GW}$.  For $\Omega_{\rm GW} \gtrsim 10^{-3}$, we reliably recover the 
input gravitational wave signal with some scatter, but a clear noise floor arises for 
$\Omega_{\rm GW} \lesssim 6\times10^{-4}$
such that the (nonsignificant) fit does not fall below $\Omega_{\rm GW} \sim 6\times10^{-4}$.  This is the limit
on the stochastic gravitational wave background that {\it Gaia} may achieve using the \citet{paine2018}
{\it Gaia}-{\it WISE} extragalactic proper motion catalog, which agrees with the rough expectation above.

\section{Discussion}\label{sec:discussion}

The VLBA+{\it Gaia} fits are less constraining on the stochastic gravitational wave background 
than the VLBA-only fits, despite the generally 
improved proper motion solutions obtained from the VLBA+{\it Gaia} time series.  Because $\Omega_{\rm GW}$
depends on the vector spherical harmonic mode power, it will scale roughly as $N$, not $\sqrt{N}$, and
the reduced VLBA+{\it Gaia} sample size should decrease sensitivity by a factor of roughly $\sqrt{2}$ compared
to the VLBA-only sample.  The sample size accounts for some, but not all, of the difference in limits on 
$\Omega_{\rm GW}$.  

The expected {\it Gaia} proper motions, despite the substantially lower precision compared to radio-based
astrometry, will further constrain the stochastic gravitational wave background by roughly an order of magnitude
due to the larger extragalactic sample
size available in visible light.  The {\it Gaia} proper motions are also expected to be less dominated by the
intrinsic proper motions caused by relativistic AGN jets.  While the VLBA+{\it Gaia} sample identified 22 new cases of
radio and optical superluminal motion and 23 cases of subluminal significant intrinsic proper motion (8\% of the 
577-object sample in total), these are radio-selected objects.  The \citet{paine2018} {\it Gaia}-{\it WISE} sample is 
optical- and infrared-selected and will therefore not be so jet-dominated as a VLBI radio-selected sample, and 
it is therefore reasonable to assume that the vast majority of {\it Gaia} objects will show no detectable 
intrinsic proper motion.  

The astrometric gravitational wave limit may be improved by increasing the sample size in either radio or visible light.
Identifying additional AGN in the {\it Gaia} catalog may decrease 
the expected $\Omega_{\rm GW}$ noise floor by at most a factor of 2 if one can achieve a $10^6$-object sample.  
In this case, one might achieve $\Omega_{\rm GW} \lesssim 2\times 10^{-4}$, which is nonetheless 
substantially larger than the $\sim10^{-6}$ value predicted by \citet{book11}.  
Improvements could also arise from an extended {\it Gaia} mission or better-than-expected performance 
of the main mission.  Enhancing the radio sample size 
is a possibility, perhaps by an order of magnitude.  Assuming similar astrometric precision and intrinsic proper 
motions to the current catalog, this would reduce the 95\% confidence limit to 
$\Omega_{\rm GW} \lesssim 6\times10^{-4}$, which is similar to the {\it Gaia} limit.

\section{Conclusions}

We have obtained limits on the low-frequency stochastic gravitational wave background using VLBA astrometry 
alone and VLBA astrometry combined with the first {\it Gaia} epoch.  We demonstrate that a quadrupole 
signal is the most constraining and obtain 95\% confidence limits on the gravitational wave energy 
density of $\Omega_{\rm GW} < 0.0064$ over the frequency range 
$6\times10^{-18}$~Hz $\lesssim f \lesssim 1\times10^{-9}$~Hz for the VLBA proper motions.
When {\it Gaia} is included, proper motion errors improve, but the limit is less constraining 
mainly due to a reduced sample size:  $\Omega_{\rm GW} < 0.011$.  The noise threshold for the VLBA fit
is roughly equivalent to a dimensionless gravitational wave strain amplitude of $h\simeq10^{-10}$ for $f\simeq10^{-9}$ Hz.

We also predict the limit that may be obtained with the full {\it Gaia} data release that includes proper motions (or limits)
of AGN.  One hurtle is the identification of AGN among the 1000-fold more numerous stars in the {\it Gaia} catalog and 
finding an all-sky distribution that is amenable to low-$\ell$ mode fitting, but provided this can be done \citep[see][]{paine2018}, 
then we predict that {\it Gaia} will find a noise floor of $\Omega_{\rm GW} \lesssim 6\times10^{-4}$ using $\sim6\times10^5$
objects.

Astrometric limits on the stochastic gravitational wave background will continue to improve with time as
geodetic monitoring of radio-loud AGN continues, but substantial improvements will need to come from 
growing the number of objects monitored.  The next post-{\it Gaia} advance could be made by a 
Next Generation Very Large Array were it to have substantial collecting area on VLBA baselines
\citep{bower2015}.

\acknowledgments

We thank David Gordon (NASA Goddard Space Flight Center) for making much of this 
work possible and Mark Reid (Smithsonian Astrophysical Observatory) for 
helpful discussions.  We also thank the anonymous referee for helpful suggestions.
The authors acknowledge support from the NSF grant AST-1411605
and the NASA grant 14-ATP14-0086.
This work has made use of data from the European Space Agency (ESA)
mission {\it Gaia} (\url{https://www.cosmos.esa.int/gaia}), processed by
the {\it Gaia} Data Processing and Analysis Consortium (DPAC,
\url{https://www.cosmos.esa.int/web/gaia/dpac/consortium}). Funding
for the DPAC has been provided by national institutions, in particular
the institutions participating in the {\it Gaia} Multilateral Agreement.
We acknowledge the 
{\it Gaia} Project Scientist Support Team and the {\it Gaia} Data Processing and Analysis Consortium
for the {\tt pyGaia} software.
This research has made use of NASA's Astrophysics Data System Bibliographic Services and
the NASA/IPAC Extragalactic Database (NED),
which is operated by the Jet Propulsion Laboratory, California Institute of Technology,
under contract with the National Aeronautics and Space Administration.

\facilities{Gaia, VLBA.}

\software{lmfit \citep{newville2014}, pyGaia.}

\appendix

\section{Vector Spherical Harmonics}\label{appendix:vsh}

Here, we present explicit formulae for the $\ell \leq 3$ vector spherical harmonics based on those described in
\citet{mignard12} for real-valued vector fields but that have strictly real coefficients (and therefore have sign differences) 
and explicitly include factors of 2 that are needed for correct power calculations.

The vector spherical harmonics for the electric and magnetic dipoles described by \citet{mignard12} and listed in Tables 
\ref{tab:Emultipoles} and \ref{tab:Bmultipoles} are 
\begin{eqnarray}
  \mathbf{\vec{V}}_{E1} (\alpha,\delta) = 
      \left(s_{11}^{Re}\, {1\over2} \sqrt{3\over\pi}\, \sin\alpha+s_{11}^{Im}\, {1\over2} \sqrt{3\over\pi}\, \cos\alpha\right) \mathbf{\hat{e}}_\alpha \nonumber\\
     +  \left(s_{10}\, {1\over2}\sqrt{3\over 2\pi}\, \cos\delta + s_{11}^{Re}\, {1\over2} \sqrt{3\over\pi}\, \cos\alpha\sin\delta
       -s_{11}^{Im}\, {1\over2} \sqrt{3\over\pi}\, \sin\alpha\sin\delta\right) \mathbf{\hat{e}}_\delta
\end{eqnarray}
and
\begin{eqnarray}
\mathbf{\vec{V}}_{M1} (\alpha,\delta) = 
      \left(t_{10}\, {1\over2}\sqrt{3\over 2\pi}\, \cos\delta + t_{11}^{Re}\, {1\over2} \sqrt{3\over\pi}\, \cos\alpha\sin\delta
       -t_{11}^{Im}\, {1\over2} \sqrt{3\over\pi}\, \sin\alpha\sin\delta\right) \mathbf{\hat{e}}_\alpha \nonumber\\
     +  \left(-t_{11}^{Re}\, {1\over2} \sqrt{3\over\pi}\, \sin\alpha-t_{11}^{Im}\, {1\over2} \sqrt{3\over\pi}\, \cos\alpha\right) \mathbf{\hat{e}}_\delta ,
\end{eqnarray}
where 
the $s_{\ell m}$ are the amplitudes of the spheroidal (curl-free or $E$-mode) orders,
the $t_{\ell m}$ are the amplitudes of the toroidal (divergence-less or $B$-mode) orders, and
$\mathbf{\hat{e}}_\alpha$ and $\mathbf{\hat{e}}_\delta$ are the unit vectors in the R.A. and decl.\ directions, respectively.

The quadrupole vector spherical harmonics are
\begin{eqnarray}
  \mathbf{\vec{V}}_{E2} (\alpha,\delta) = 
      \left(s_{21}^{Re}\, {1\over2} \sqrt{5\over\pi}\, \sin\alpha\sin\delta+s_{21}^{Im}\, {1\over2} \sqrt{5\over\pi}\, \cos\alpha\sin\delta
- s_{22}^{Re}\, {1\over2} \sqrt{5\over\pi}\, \sin2\alpha\cos\delta \right.\nonumber\\ 
       \left. -s_{22}^{Im}\, {1\over2} \sqrt{5\over\pi}\, \cos2\alpha\cos\delta\right) \mathbf{\hat{e}}_\alpha \nonumber\\
     + \left(s_{20}\, {1\over4}\sqrt{15\over 2\pi}\, \sin2\delta - s_{21}^{Re}\, {1\over2} \sqrt{5\over\pi}\, \cos\alpha\cos2\delta
     + s_{21}^{Im}\, {1\over2} \sqrt{5\over\pi}\, \sin\alpha\cos2\delta \right.\nonumber\\
       \left. - s_{22}^{Re}\, {1\over4}\sqrt{5\over\pi}\, \cos2\alpha\sin2\delta 
      + s_{22}^{Im}\, {1\over4}\sqrt{5\over\pi}\, \sin2\alpha\sin2\delta \right) \mathbf{\hat{e}}_\delta  
  \label{eqn:Equad}
\end{eqnarray}
and
\begin{eqnarray}
  \mathbf{\vec{V}}_{M2} (\alpha,\delta) = 
     \left(t_{20}\, {1\over4}\sqrt{15\over 2\pi}\, \sin2\delta - t_{21}^{Re}\, {1\over2} \sqrt{5\over\pi}\, \cos\alpha\cos2\delta
     + t_{21}^{Im}\, {1\over2} \sqrt{5\over\pi}\, \sin\alpha\cos2\delta \right.\nonumber\\
       \left. - t_{22}^{Re}\, {1\over4}\sqrt{5\over\pi}\, \cos2\alpha\sin2\delta 
      + t_{22}^{Im}\, {1\over4}\sqrt{5\over\pi}\, \sin2\alpha\sin2\delta \right) \mathbf{\hat{e}}_\alpha \nonumber\\
      +\left(-t_{21}^{Re}\, {1\over2} \sqrt{5\over\pi}\, \sin\alpha\sin\delta-t_{21}^{Im}\, {1\over2} \sqrt{5\over\pi}\, \cos\alpha\sin\delta
      + t_{22}^{Re}\, {1\over2} \sqrt{5\over\pi}\, \sin2\alpha\cos\delta \right.\nonumber\\ 
       \left. + t_{22}^{Im}\, {1\over2} \sqrt{5\over\pi}\, \cos2\alpha\cos\delta\right) \mathbf{\hat{e}}_\delta .
\end{eqnarray}
The octopole ($\ell=3$) vector spherical harmonics are 
\begin{eqnarray}
  \mathbf{\vec{V}}_{E3} (\alpha,\delta) = 
      \left(s_{31}^{Re}\, {1\over8} \sqrt{7\over\pi}\, \sin\alpha(5\sin^2\delta-1)
              + s_{31}^{Im}\, {1\over8} \sqrt{7\over\pi}\, \cos\alpha(5\sin^2\delta-1) \right. \nonumber\\
    \left.  -\ s_{32}^{Re}\, {1\over4} \sqrt{35\over2\pi}\, \sin2\alpha\sin2\delta 
             - s_{32}^{Im}\, {1\over4} \sqrt{35\over2\pi}\, \cos2\alpha\sin2\delta 
             + s_{33}^{Re}\, {1\over8} \sqrt{105\over\pi}\, \sin3\alpha\cos^2\delta \right.\nonumber\\
   \left.  +\ s_{33}^{Im}\, {1\over8} \sqrt{105\over\pi}\, \cos3\alpha\cos^2\delta \right) \mathbf{\hat{e}}_\alpha \nonumber\\
    + \left(s_{30}\, {1\over8} \sqrt{21\over\pi}\, (5\sin^2\delta-1)\cos\delta
            + s_{31}^{Re}\, {1\over8} \sqrt{7\over\pi}\, \cos\alpha\sin\delta(15\sin^2\delta-11)  \right. \nonumber\\
   \left. -\ s_{31}^{Im}\, {1\over8} \sqrt{7\over\pi}\, \sin\alpha\sin\delta(15\sin^2\delta-11)  
            - s_{32}^{Re}\, {1\over4} \sqrt{35\over2\pi}\, \cos2\alpha\cos\delta(3\sin^2\delta-1) \right. \nonumber\\
   \left.   +\ s_{32}^{Im}\, {1\over4} \sqrt{35\over2\pi}\, \sin2\alpha\cos\delta(3\sin^2\delta-1) 
    + s_{33}^{Re}\, {1\over8} \sqrt{105\over\pi}\, \cos3\alpha\cos^2\delta\sin\delta \right. \nonumber\\
    \left.  -\ s_{33}^{Im}\, {1\over8} \sqrt{105\over\pi}\, \sin3\alpha\cos^2\delta\sin\delta \right) \mathbf{\hat{e}}_\delta 
\end{eqnarray}
and 
\begin{eqnarray}
  \mathbf{\vec{V}}_{M3} (\alpha,\delta) = 
      \left(t_{30}\, {1\over8} \sqrt{21\over\pi}\, (5\sin^2\delta-1)\cos\delta 
             +  t_{31}^{Re}\, {1\over8} \sqrt{7\over\pi}\, \cos\alpha\sin\delta(15\sin^2\delta-11)  \right. \nonumber\\   \left.
             -\ t_{31}^{Im}\, {1\over8} \sqrt{7\over\pi}\, \sin\alpha\sin\delta(15\sin^2\delta-11)  
             - t_{32}^{Re}\, {1\over4} \sqrt{35\over2\pi}\, \cos2\alpha\cos\delta(3\sin^2\delta-1) \right. \nonumber\\ \left.
            +\ t_{32}^{Im}\, {1\over4} \sqrt{35\over2\pi}\, \sin2\alpha\cos\delta(3\sin^2\delta-1) 
             + t_{33}^{Re}\, {1\over8} \sqrt{105\over\pi}\, \cos3\alpha\cos^2\delta\sin\delta \right. \nonumber\\ \left.
     -\ t_{33}^{Im}\, {1\over8} \sqrt{105\over\pi}\, \sin3\alpha\cos^2\delta\sin\delta \right) \mathbf{\hat{e}}_\alpha \nonumber\\
    + \left(-\ t_{31}^{Re}\, {1\over8} \sqrt{7\over\pi}\, \sin\alpha(5\sin^2\delta-1)  
  -\ t_{31}^{Im}\, {1\over8} \sqrt{7\over\pi}\, \cos\alpha(5\sin^2\delta-1)  \right. \nonumber\\  \left.
            +\ t_{32}^{Re}\, {1\over4} \sqrt{35\over2\pi}\, \sin2\alpha\sin2\delta 
      +\ t_{32}^{Im}\, {1\over4} \sqrt{35\over2\pi}\, \cos2\alpha\sin2\delta \right. \nonumber\\ \left.
    -\ t_{33}^{Re}\, {1\over8} \sqrt{105\over\pi}\, \sin3\alpha\cos^2\delta 
    -\ t_{33}^{Im}\, {1\over8} \sqrt{105\over\pi}\, \cos3\alpha\cos^2\delta \right) \mathbf{\hat{e}}_\delta .
  \label{eqn:Boct}
\end{eqnarray}

\begin{deluxetable}{lccc}[h]
\tabletypesize{\scriptsize} 
\tablecaption{Spheroidal ($E$-mode) Vector Spherical Harmonics ($\ell\leq3$) \label{tab:Emultipoles}}
\tablewidth{0pt} \tablehead{
\colhead{$\ell,m$} &
\colhead{Amplitude} &
\colhead{$\mathbf{\hat{e}}_\alpha$} &
\colhead{$\mathbf{\hat{e}}_\delta$}}
\startdata 
1,0       & ${1\over2}\sqrt{3\over2\pi}$ & 0 & $\cos\delta$ \\
1,1 (Re) & ${1\over2}\sqrt{3\over\pi}$ & $\sin\alpha$ & $\cos\alpha\sin\delta$ \\
1,1 (Im) & ${1\over2}\sqrt{3\over\pi}$ & $\cos\alpha$ & $-\sin\alpha\sin\delta$ \\
2,0       & ${1\over4}\sqrt{15\over2\pi}$ & 0 & $\sin2\delta$ \\
2,1 (Re) & ${1\over2}\sqrt{5\over\pi}$ & $\sin\alpha\sin\delta$ & $-\cos\alpha\cos2\delta$ \\
2,1 (Im) & ${1\over2}\sqrt{5\over\pi}$ & $\cos\alpha\sin\delta$ & $\sin\alpha\cos2\delta$ \\
2,2 (Re) & ${1\over4}\sqrt{5\over\pi}$ & $-2\sin2\alpha\cos\delta$ & $-\cos2\alpha\sin2\delta$ \\
2,2 (Im) & ${1\over4}\sqrt{5\over\pi}$ & $-2\cos2\alpha\cos\delta$ & $\sin2\alpha\sin2\delta$ \\
3,0       & ${1\over8}\sqrt{21\over\pi}$ & 0 & $(5\sin^2\delta-1)\cos\delta$ \\
3,1 (Re) & ${1\over8}\sqrt{7\over\pi}$ & $\sin\alpha(5\sin^2\delta-1)$ & $\cos\alpha\sin\delta(15\sin^2\delta-11)$ \\
3,1 (Im) & ${1\over8}\sqrt{7\over\pi}$ & $\cos\alpha(5\sin^2\delta-1)$ & $-\sin\alpha\sin\delta(15\sin^2\delta-11)$\\ 
3,2 (Re) & ${1\over4}\sqrt{35\over2\pi}$ & $-\sin2\alpha\sin2\delta$ & $-\cos2\alpha\cos\delta(3\sin^2\delta-1)$ \\
3,2 (Im) & ${1\over4}\sqrt{35\over2\pi}$ & $-\cos2\alpha\sin2\delta$ & $\sin2\alpha\cos\delta(3\sin^2\delta-1)$ \\
3,3 (Re) & ${1\over8}\sqrt{105\over\pi}$ & $\sin3\alpha\cos^2\delta$ & $\cos3\alpha\cos^2\delta\sin\delta$ \\
3,3 (Im) & ${1\over8}\sqrt{105\over\pi}$ & $\cos3\alpha\cos^2\delta$ & $-\sin3\alpha\cos^2\delta\sin\delta$ \\
\enddata 
\end{deluxetable}

\begin{deluxetable}{lccc}
\tabletypesize{\scriptsize} 
\tablecaption{Toroidal ($B$-mode) Vector Spherical Harmonics ($\ell\leq3$) \label{tab:Bmultipoles}}
\tablewidth{0pt} \tablehead{
\colhead{$\ell,m$} &
\colhead{Amplitude} &
\colhead{$\mathbf{\hat{e}}_\alpha$} &
\colhead{$\mathbf{\hat{e}}_\delta$}}
\startdata 
1,0       & ${1\over2}\sqrt{3\over2\pi}$ & $\cos\delta$ & 0 \\ 
1,1 (Re) & ${1\over2}\sqrt{3\over\pi}$ & $\cos\alpha\sin\delta$ &$-\sin\alpha$ \\
1,1 (Im) & ${1\over2}\sqrt{3\over\pi}$ & $-\sin\alpha\sin\delta$ & $-\cos\alpha$ \\
2,0       & ${1\over4}\sqrt{15\over2\pi}$ & $\sin2\delta$ & 0 \\
2,1 (Re) & ${1\over2}\sqrt{5\over\pi}$ & $-\cos\alpha\cos2\delta$ &$-\sin\alpha\sin\delta$ \\
2,1 (Im) & ${1\over2}\sqrt{5\over\pi}$ & $\sin\alpha\cos2\delta$ &$-\cos\alpha\sin\delta$ \\
2,2 (Re) & ${1\over4}\sqrt{5\over\pi}$ & $-\cos2\alpha\sin2\delta$ &$2\sin2\alpha\cos\delta$ \\
2,2 (Im) & ${1\over4}\sqrt{5\over\pi}$ & $\sin2\alpha\sin2\delta$ &$2\cos2\alpha\cos\delta$ \\
3,0       & ${1\over8}\sqrt{21\over\pi}$ & $(5\sin^2\delta-1)\cos\delta$ & 0 \\
3,1 (Re) & ${1\over8}\sqrt{7\over\pi}$ & $\cos\alpha\sin\delta(15\sin^2\delta-11)$ &$-\sin\alpha(5\sin^2\delta-1)$ \\
3,1 (Im) & ${1\over8}\sqrt{7\over\pi}$ & $-\sin\alpha\sin\delta(15\sin^2\delta-11)$ &$-\cos\alpha(5\sin^2\delta-1)$ \\
3,2 (Re) & ${1\over4}\sqrt{35\over2\pi}$ & $-\cos2\alpha\cos\delta(3\sin^2\delta-1)$ &$\sin2\alpha\sin2\delta$ \\
3,2 (Im) & ${1\over4}\sqrt{35\over2\pi}$ & $\sin2\alpha\cos\delta(3\sin^2\delta-1)$ &$\cos2\alpha\sin2\delta$ \\
3,3 (Re) & ${1\over8}\sqrt{105\over\pi}$ & $\cos3\alpha\cos^2\delta\sin\delta$ &$-\sin3\alpha\cos^2\delta$ \\
3,3 (Im) & ${1\over8}\sqrt{105\over\pi}$ & $-\sin3\alpha\cos^2\delta\sin\delta$ &$-\cos3\alpha\cos^2\delta$ \\
\enddata 
\end{deluxetable}


\begin{deluxetable*}{crrcccccc}
\tabletypesize{\scriptsize} 
\tablecaption{Objects Showing Significant Consistent Radio and Optical Proper Motion\label{tab:superluminal}}
\tablewidth{0pt} 
\tablehead{
\colhead{Name} &
\multicolumn{2}{c}{VLBA+Gaia PM} &
\colhead{Redshift} & 
\colhead{References\tablenotemark{a}} & 
\colhead{$D_M$} & 
\multicolumn{3}{c}{Apparent Velocity} \\
\cline{2-3} \cline{7-9}
\colhead{} &
\colhead{$\mu_\alpha$} & 
\colhead{$\mu_\delta$} &
\colhead{} &
\colhead{} &
\colhead{} &
\colhead{$v_\alpha$} & 
\colhead{$v_\delta$} &
\colhead{$v_{\rm Total}$} \\
\colhead{} & 
\colhead{($\mu$as yr$^{-1}$)} & 
\colhead{($\mu$as yr$^{-1}$)} & 
\colhead{} &
\colhead{} &
\colhead{(Mpc)} &
\colhead{($c$)} & 
\colhead{($c$)} & 
\colhead{($c$)}}
\startdata 
{\bf 0007+171} & 136.6(25.4) & $-$0.3(12.4) & 1.60 & 1 & 4654 & {\bf 10.1(1.9)} & $-$0.02(0.91) & {\bf 10.1(1.9)} \\
0016+731 & $-$5.5(1.1) & 5.1(0.7) & 1.78 & 2 & 4969 &  $-$0.43(0.09) & 0.40(0.06) & 0.59(0.07)\\
0059+581 & $-$7.5(0.4) & $-$2.8(0.5) & 0.64 & 3 & 2359 & $-$0.28(0.01) & $-$0.10(0.02) & 0.30(0.02) \\
0119+041 & $-$9.2(1.8) & 6.0(1.9) & 0.64 & 2 & 2359 & $-$0.34(0.07) & 0.22(0.07) & 0.41(0.07) \\
{\bf 0229+131} & 10.7(0.7) & 7.0(0.7) & 2.06 & 2 & 5409 & 0.92(0.06) & 0.60(0.06) &  {\bf 1.09(0.06)} \\
 {\bf NRAO150} & 6.0(1.9) & 16.6(2.3) & 1.52 & 4 & 4505 & 0.43(0.14) &  {\bf 1.18(0.16)} & {\bf 1.26(0.16)} \\
0420$-$014 & $-$5.5(0.7) & $-$8.5(0.9) & 0.92 & 2 & 3158 & $-$0.27(0.03) & $-$0.42(0.04) & 0.51(0.04) \\
{\bf NRAO190} & $-$20.6(9.9) & $-$28.6(3.3) & 0.84 & 2 & 2943 & $-$0.96(0.46) &  {\bf $-$1.33(0.15)} &  {\bf 1.64(0.30)}\\ 
0454$-$234 &  $-$3.8(0.7) & $-$7.1(0.8) & 1.00 & 2 & 3365 & $-$0.20(0.04) & $-$0.38(0.04) & 0.43(0.04) \\ 
0458$-$020 &  $-$2.5(0.9) & $-$10.4(0.8) & 2.29 & 2 & 5731 & $-$0.23(0.08) & $-$0.94(0.07) & 0.97(0.07) \\ 
{\bf 0454+844} & 6.9(4.7) & 19.7(3.0) & 1.34 & 2 & 4145 & 0.45(0.31) &  {\bf 1.29(0.20)} &  {\bf 1.37(0.21)} \\  
0552+398 & 0.2(0.4) & $-$3.5(0.4) & 2.37 & 1 & 5835 & 0.02(0.04) & $-$0.32(0.04) & 0.32(0.04) \\ 
{\bf 0602+673} &  0.5(0.8) & 20.4(1.2) & 1.97 & 2 & 5274 & 0.04(0.07) &  {\bf 1.70(0.10)} &  {\bf 1.70(0.10)} \\ 
0657+172 & 7.2(1.4) & $-$5.4(1.9) & 1.08 & 5 & 3562 & 0.41(0.08) & $-$0.30(0.11) & 0.51(0.09) \\ 
{\bf 0723$-$008} & $-$50.0(6.5) & 107.5(13.1) & 0.13 & 1 & 542 & $-$0.43(0.06) & 0.92(0.11) & {\bf 1.02(0.10)} \\ 
{\bf 0743+259} & $-$3.5(2.8) & $-$33.4(3.6) & 2.99 & 6 & 6543 & $-$0.36(0.29) & {\bf $-$3.46(0.37)} & {\bf 3.47(0.37)}\\ 
0805+410 &  5.4(1.1) & 9.4(1.3) & 1.42 & 2 & 4309 & 0.37(0.07) & 0.64(0.09) & 0.74(0.09) \\ 
{\bf 1038+064} & $-$15.1(3.5) & 61.6(8.4) & 1.27 & 2 & 3996 & $-$0.95(0.22) & {\bf 3.89(0.53)} & {\bf 4.01(0.52)} \\ 
{\bf 1045$-$188} & 22.3(4.3) & $-$73.5(10.4) & 0.59 & 2 & 2202 & 0.78(0.15) & {\bf $-$2.56(0.36)} & {\bf 2.67(0.35)} \\ 
1053+815 & $-$7.1(1.4) & 1.8(1.5) & 0.71 & 3 & 2571 & $-$0.29(0.06) & 0.07(0.06) & 0.30(0.06)\\ 
1057$-$797 & 0.5(1.3) & $-$7.5(1.2) & 0.58 & 7 & 2170 & 0.02(0.04) & $-$0.26(0.04) & 0.26(0.04) \\ 
{\bf 1104$-$445} & $-$20.1(3.0) & 12.4(3.2) & 1.60 & 2 & 4654 & {\bf $-$1.48(0.22)} & 0.91(0.24) & {\bf 1.74(0.22)} \\ 
1124$-$186 & 1.7(1.1) & $-$6.8(1.2) & 1.05 & 2 & 3489 & 0.09(0.06) & $-$0.38(0.07) & 0.39(0.07) \\ 
1219+044 & 6.3(1.1) & $-$3.6(2.0) & 0.97 & 2 & 3288 & 0.33(0.06) & $-$0.19(0.10) & 0.38(0.07) \\ 
1300+580 & 5.2(0.7) & 11.4(0.8) & 1.09 & 2 & 3586 & 0.29(0.04) & 0.65(0.05) & 0.71(0.04) \\ 
{\bf 1342+663} &  $-$40.9(7.3) & $-$9.8(2.8) & 1.35 & 2 & 4166 & {\bf $-$2.69(0.48)} & $-$0.65(0.18) & {\bf 2.77(0.47)} \\ 
1424$-$418 & $-$9.8(1.6) & 2.0(1.8) & 1.52 & 2 & 4505 & $-$0.70(0.11) & 0.14(0.13) & 0.71(0.11) \\ 
1606+106 & 5.4(1.0) & 0.1(0.9) & 1.23 & 2 & 3908 & 0.32(0.06) & 0.01(0.06) & 0.32(0.06) \\ 
1622$-$253 & $-$0.2(1.1) & 8.0(1.4) & 0.79 & 2 & 2803 & $-$0.01(0.05) & 0.35(0.06) & 0.35(0.06) \\ 
1642+690 & 6.3(1.6) & $-$19.3(3.0) & 0.75 & 2 & 2688 & 0.27(0.07) & $-$0.82(0.13) & 0.86(0.12) \\ 
1657$-$562\tablenotemark{b}& 33.7(10.1) & $-$110.7(17.0) & \nodata & \nodata & \nodata & \nodata & \nodata & \nodata \\ 
NRAO530  &  7.4(1.3) & 7.2(2.3) & 0.90 & 2 & 3105 & 0.36(0.06) & 0.35(0.11) & 0.51(0.09) \\ 
{\bf 1745+624} & 10.7(1.5) & 10.4(2.2) & 3.89 & 8 &  7328 & {\bf 1.24(0.17)} & {\bf 1.21(0.25)} & {\bf 1.73(0.22)} \\ 
{\bf 1846+322} & $-$29.5(4.7) & 7.2(5.9) & 0.80 & 2 & 2831 & {\bf $-$1.32(0.21)} & 0.32(0.26) & {\bf 1.36(0.21)} \\ 
{\bf 3C395} &  66.2(9.9) & $-$32.1(6.9) & 0.64 & 9 & 2359 & {\bf 2.47(0.37)} & {\bf $-$1.20(0.26)} & {\bf 2.74(0.35)} \\ 
1923+210\tablenotemark{b} & 10.0(2.4) & 17.9(1.4) & \nodata & \nodata & \nodata & \nodata & \nodata & \nodata \\ 
1958$-$179 &  $-$7.7(1.0) & $-$4.0(1.4) & 0.65 & 2 & 2390 & $-$0.29(0.04) & $-$0.15(0.05) & 0.33(0.04) \\ 
2007+777 & 22.8(3.0) & $-$0.0(1.5) & 0.34 & 2 & 1350 & 0.49(0.06) & $-$0.00(0.03) & 0.49(0.06) \\ 
{\bf 3C418} & $-$15.3(1.2) & $-$7.7(1.6) & 1.69 & 10 & 4815 & {\bf $-$1.16(0.09)} & $-$0.59(0.12) & {\bf 1.30(0.10)} \\  
{\bf 2059+034} & $-$5.3(3.5) & $-$25.0(3.8) & 1.01 & 2 & 3390 & $-$0.28(0.19) & {\bf $-$1.34(0.20)} & {\bf 1.37(0.20)} \\ 
{\bf 2126$-$158} & $-$10.6(2.5) & $-$66.5(5.2) &  3.27 & 2 & 6812 & {\bf $-$1.14(0.27)} & {\bf $-$7.16(0.56)} & {\bf 7.25(0.55)} \\ 
{\bf 2155$-$152} & $-$30.4(6.3) & $-$40.1(7.3) & 0.67 & 2 & 2451 & {\bf $-$1.18(0.24)} & {\bf $-$1.55(0.28)} & {\bf 1.95(0.27)} \\ 
{\bf 2209+236} & 21.7(2.0) & 2.6(2.0) & 1.13 & 2 &  3680 & {\bf 1.26(0.12)} & 0.15(0.12) & {\bf 1.27(0.12)} \\ 
{\bf 2214+350} & 5.6(3.1) & $-$69.5(5.0) & 0.51 & 1 & 1942 & 0.17(0.10) & {\bf $-$2.13(0.15)} & {\bf 2.14(0.15)} \\ 
{\bf 2229+695} & 39.3(3.3) & 4.8(1.6) & 1.41 & 2 & 4289 & {\bf 2.67(0.22)} & 0.33(0.11) & {\bf 2.69(0.22)} \\ 
\enddata
\tablecomments{Parenthetical values are 1$\sigma$ uncertainties.  Apparent velocities are in the rest frame of each 
object, in units of the speed of light, $c$ (see Section \ref{subsec:vlba-gaia}).   
Bold type indicates the names and velocities of objects showing superluminal motion.}
\tablenotetext{a}{Redshift references:  
(1) \citet{wills1976}; 
(2) \citet{healey2008};
(3) \citet{sowards-emmerd2005};
(4) \citet{agudo2007};
(5) \citet{alvarezcrespo2016};
(6) \citet{hewett2010};
(7) \citet{sbarufatti2009};
(8) \citet{hook1995};
(9) \citet{gelderman1994};
(10) \citet{smith1980}.
}
\tablenotetext{b}{1657$-$562 and 1923+210 show VLBA+{\it Gaia} astrometric correspondence and significant 
proper motion, but the redshifts for these objects are unknown.  If they have redshifts greater than 0.13 and 0.90, 
respectively, then their observed proper motions would be superluminal.}
\end{deluxetable*}

\clearpage 

\section{VLBA+{\it Gaia} Astrometry and Proper Motions}\label{appendix:pms}

Table \ref{tab:superluminal} lists the objects that show significant and consistent radio and optical 
proper motion.  Based on the amplitudes of the proper motions, these are most likely intrinsic and 
associated with jets.  Objects showing apparent superluminal motions are indicated in bold, as are the
superluminal velocity components and amplitudes.  

Table \ref{tab:pms} lists the {\it Gaia} 2015.0 epoch J2000 coordinates, the VLBA-only and VLBA+{\it Gaia} proper
motions obtained from the time series fits described in Section \ref{subsec:vlba-gaia}, and the {\it Gaia}-VLBA 
coordinate offset in {\it Gaia} standard deviations.  
The VLBA proper motions were obtained from bootstrap-resampled time series 
and may differ slightly from --- but are statistically consistent with --- the \citet{truebenbach2017} 
proper motion catalog.  The VLBA+{\it Gaia} proper motion catalog in Table \ref{tab:pms} 
forms a subset of the VLBA-only
\citet{truebenbach2017} catalog because not all radio sources have {\it Gaia} counterparts.

\startlongtable 
\begin{deluxetable*}{cllrrrrcc}
\tabletypesize{\scriptsize} 
\tablecaption{VLBA+{\it Gaia} Astrometry and Proper Motions\label{tab:pms}}
\tablewidth{0pt} 
\tablehead{
\colhead{Name} &
\multicolumn{2}{c}{Gaia 2015.0 Coordinates} &
\multicolumn{2}{c}{VLBA PM} &
\multicolumn{2}{c}{VLBA+Gaia PM} &
\multicolumn{2}{c}{{\it Gaia}-VLBA Offset} \\
\colhead{} &
\colhead{R.A. (J2000)} & 
\colhead{Decl.\ (J2000)} & 
\colhead{$\mu_\alpha$} & 
\colhead{$\mu_\delta$} &
\colhead{$\mu_\alpha$} & 
\colhead{$\mu_\delta$} &
\colhead{$\Delta\alpha$} &
\colhead{$\Delta\delta$} \\
\colhead{} & 
\colhead{(h:m:s)} & 
\colhead{(d:m:s)} & 
\colhead{($\mu$as yr$^{-1}$)} & 
\colhead{($\mu$as yr$^{-1}$)} & 
\colhead{($\mu$as yr$^{-1}$)} & 
\colhead{($\mu$as yr$^{-1}$)} & 
\colhead{($\sigma$)} &
\colhead{($\sigma$)}}
\startdata 
0002$-$478 & 00:04:35.65554(6) & $-$47:36:19.6040(8) & $-$23.1(13.2) & $-$46.5(20.0) & $-$22.9(13.2) & $-$47.4(19.8) & 1.1 & 0.2 \\
0003+380 & 00:05:57.17593(8) & +38:20:15.1435(4) & $-$15.4(9.9) & $-$1.5(11.3) & $-$12.6(9.2) & $-$18.9(13.6) & 6.7 & 14.2 \\
0003$-$066 & 00:06:13.89290(2) & $-$06:23:35.3351(1) & 0.3(1.3) & 3.7(1.8) & 0.4(1.3) & 3.5(1.8) & 0.4 & 1.4 \\
IIIZW2 & 00:10:31.00592(2) & +10:58:29.5038(2) & 2.3(12.3) & $-$9.7(11.8) & 3.3(11.5) & $-$13.1(12.0) & 0.9 & 2.2 \\
0007+171 & 00:10:33.99072(4) & +17:24:18.7609(3) & 136.2(27.2) & 9.8(14.7) & 136.6(25.4) & $-$0.3(12.4) & 0.6 & 2.0 \\
0008$-$264 & 00:11:01.24678(3) & $-$26:12:33.3768(3) & 1.1(6.8) & $-$0.4(8.2) & 1.2(6.9) & $-$1.5(8.3) & 1.3 & 0.8 \\
0010+405 & 00:13:31.13019(11) & +40:51:37.1457(6) & $-$3.9(3.1) & 7.2(4.6) & $-$3.6(3.0) & 7.7(4.7) & 0.1 & 2.6 \\
0013$-$005 & 00:16:11.08856(9) & $-$00:15:12.4450(7) & 1.8(5.3) & 3.6(7.3) & 1.6(4.8) & 3.3(7.8) & 0.0 & 0.5 \\
\enddata 
\tablecomments{The {\it Gaia} coordinates are for epoch 2015.0, expressed in the J2000 reference frame.  
The {\it Gaia}-VLBA offsets are expressed in standard deviations of the {\it Gaia} astrometry from the
 VLBA time series fit, assuming that the error is dominated by the sole {\it Gaia} epoch in each coordinate.}
\end{deluxetable*}

.

\end{document}